\documentclass[usenatbib,usegraphicx]{mn2e}

\title[
Effects of a deformation of a star]
{
Effects of a deformation of a star on the gravitational lensing
}
\author[H. Asada]
  {H.~Asada\thanks{Email: asada@phys.hirosaki-u.ac.jp}\\ 
GReCO, Institute for Astrophysics at Paris, 
98bis boulevard Arago, 75014 Paris, France\\
Faculty of Science and Technology, 
Hirosaki University, Hirosaki 036-8561, Japan}

\begin{document}

\date{Accepted  Received }

\pagerange{\pageref{firstpage}--\pageref{lastpage}}

\pubyear{2004}

\maketitle

\label{firstpage}

\begin{abstract}
We study analytically a gravitational lens due to a deformed star, 
which is modeled by using a monopole and a quadrupole moment. 
Positions of the images are discussed for a source 
on the principal axis. 
We present explicit expressions for the lens equation 
for this gravitational lens as a single real tenth-order 
algebraic equation.  
Furthermore, we compute an expression for the caustics 
as a discriminant for the polynomial. 
Another simple parametric representation of the caustics 
is also presented in a more tractable form. 
A simple expression for the critical curves is obtained 
to clarify a topological feature of the critical curves; 
the curves are simply connected if and only if the distortion 
is sufficiently large. 
\end{abstract}


\section{Introduction}
Some recent highly accurate numerical simulations of 
rotating compact objects such as a neutron star 
elucidate many features of compact objects such as 
an equilibrium configuration and its stability 
(e.g., Shibata et al. 2002, Stergioulas et al. 2004, 
Zanotti and Rezzolla 2002, Zdunik et al. 2004). 
The deformation of their configurations causes 
a non-spherical distortion of the surrounding spacetime, 
which has not been observed yet. 
Probing this warp of the spacetime provides a strong verification 
of the theory of the general relativity (Will 1993) and 
a deep insight into the relevant physics especially 
about the nuclear matter (e.g., Shapiro and Teukolsky 1983), 
because the deformation depends not only on the gravity 
but also on the equation of state of the internal matter. 

One possibility is to study effects of the deformation 
on the light propagation near the compact object. 
Before doing numerical studies in detail, it is important 
to find some physical properties by using a simple analytic model. 
One reason is that one needs many light ray tracings in 
numerical investigations, and consequently it is difficult 
to clarify quantitatively the parameter dependence. 

We perform the multipole expansion of the external gravitational 
field, which can be assumed to be weak along the line of sight 
even if the internal gravity is strong. 
Namely, one can consider only the linearized potential. 
This does not mean that our system is linear, because 
the lens equation is non-linearly coupled with respect to 
the image positions as shown in the next section.  

The quadrupole moment depends on the rotation speed and the equation 
of the state for the internal matter. 
Many years ago some comprehensive studies had been done for 
equilibrium configurations of rotating stars in the Newtonian gravity, 
some of which are summarized in a famous textbook (Chandrasekhar 1987). 
Numerical investigations on this issue in the general relativity 
are still going on as mentioned in the beginning of this section 
(e.g., Shibata et al. 2002, Stergioulas et al. 2004, 
Zanotti and Rezzolla 2002, Zdunik et al. 2004).  
In an extreme case called as rapid rotation, 
the quadrupole deviation of order of $10^{-3}$ may be possible 
even for a neutron star. 
These results show that the deformation is dominated by the 
quadrupole moment and higher moments such as the octapole 
moment are much smaller. 
In addition, a contribution of the octapole moment in the 
external field decays faster than that of the quadrupole moment, 
because the ratio between the octapole potential and the quadrupole 
potential is of the order of the ratio of the stellar radius 
to the distance between the star and the field point, which 
becomes the impact parameter in the gravitational lensing.  
This ratio is so small that one can ignore approximately 
the octapole and higher moment in the lensing study. 
Furthermore, radio observations such as the VLBI will be improved. 
If the accuracy is improved sufficiently, the present result 
will have a direct relevance with such an observation. 
Hence, we consider in this paper the truncated multipole 
expansion up to the quadrupole moment. 

This consideration in the gravitational lens is not new 
but rather classical though 
it was done in a different context. 
For instance, the light deflection caused by the quadrupole 
moment of the Sun was discussed in the context of 
tests of the general relativity (Epstein and Shapiro 1980). 
Except for a case of the source on the equatorial plane, 
we need solve the lens equation numerically. 
A spherically symmetric galaxy and a quadrupole tidal part 
were extensively studied by Kovner (1987), in which 
most models are an extended sphere and the quadrupole part, 
and another model is a star with quadrupole shear induced by 
a galaxy where the star is located. The last model seems similar 
to our system of a deformed star. However, it should be noted that 
this extra shear does not come from a potential of the star 
but from that of the galaxy. 
In other words, the dependence of the shear on 
the radial coordinate is different from that for the deformed star. 

It is believed that gravitational lensing study needs 
some numerical techniques except for a few special cases 
such as a point mass lens or an axisymmetric mass distribution 
along the line of sight, because the lens equation is nonlinearly 
coupled (for instance, Schneider et al. 1992). 
In recent, however, some development has been achieved: 
It has been shown that for two cases 
of a binary lens (Asada 2002) and 
an ellipsoidal lens (Asada et al. 2003) 
the lens equation is reduced to a single polynomial. 
An advantage of this approach is saving time and computer resources, 
and another one is to enable us to study analytically the lensing, 
for instance to find out an analytic expression 
for caustics (Asada et al. 2002, 2003) which are curves 
on the source plane, and the number of images changes 
if a source crosses the caustics. 
This expression for the caustics is nothing but a discriminant 
for the polynomial, which determines a change in the number of roots 
for the equation: 
As known in algebra, the discriminant vanishes in a case of 
multiple roots, which corresponds to merging images or giant arcs 
in the gravitational lens. Classification of the critical curves and 
caustics is of great importance. For instance, it was done intensively 
for a binary gravitational lens (e.g., Dominik 1999). 

Furthermore, it has been shown also that one can derive 
a single polynomial equation for any set of the lens equations 
in polynomials by using the Euclidean algorithm (Asada et al. 2004). 
In practice, however, in a case of the lens equation 
which is higher than fifth degree, this algebraic procedure 
needs a computer with a fast CPU and a huge memory, 
and furthermore the resultant equation may become too complicated 
to handle any more in the Cartesian coordinates. In order to avoid 
this problem, the polar coordinates are adopted in this paper 
so that a simple result can be obtained. 

This paper is organised as follows. 
First, we show that the lens equation is reduced to a single 
polynomial equation. Next, the analytic expression for the caustics 
is computed as the discriminant for the polynomial. 
Finally, a parametric representation for the critical curves 
and caustics are obtained to clarify a topological feature 
of the critical curves.

\section{Lens Equation for a Lens with Quadrupole Moment} 
\subsection{Lens Equation}
The lens equation relates the image angular position vector  
$\mbox{\boldmath $\theta$}$ 
to the source angular position vector 
 $\mbox{\boldmath $\beta$}$ 
as 
\begin{equation}
  \mbox{\boldmath $\beta$} = \mbox{\boldmath $\theta$} 
- \frac{D_{\mbox{LS}}}{D_{\mbox{S}}}
\mbox{\boldmath $\alpha$}(\mbox{\boldmath $\xi$}),   
\label{lenseq1} 
\end{equation}
where we used the thin-lens approximation, 
$D_{\mbox{S}}$ is the angular distance from the observer 
to the source, and $D_{\mbox{LS}}$ is that from the lens 
to the source. 
The vector $\mbox{\boldmath $\xi$}$, which denotes 
the impact parameter to the light ray, is related 
to the image position as 
$\mbox{\boldmath $\xi$}=D_{\mbox{L}} \mbox{\boldmath $\theta$}$, 
where $D_{\mbox{L}}$ is the angular distance from the observer 
to the lens.  

Up to quadrupole moments in multipole expansions, 
the deflection angle 
$\mbox{\boldmath $\alpha$} = (\alpha^1, \alpha^2)$ 
is given by 
\begin{equation}
  \alpha^i = \frac{4GM}{c^2} \frac{\xi^i}{|\mbox{\boldmath $\xi$}|^2}
  +\frac{8G}{c^2}\Bigl(
\sum_{j,k=1}^{2} 2{Q}_{jk} 
\frac{\xi^j \xi^k \xi^i}{|\mbox{\boldmath $\xi$}|^6}
- \sum_{j=1}^{2} {Q}_{ij}\frac{\xi^j}{|\mbox{\boldmath $\xi$}|^4} 
\Bigr) . 
\label{deflection1}
\end{equation}
Here $M$ denotes the mass of the lens, and 
the quadrupole moment tracefree on the lens plane 
is defined as 
\begin{equation}
Q_{ij}=\int\rho (X_i X_j - \frac12 \delta_{ij} 
|\mbox{\boldmath $X$}|^2) d^3X , 
\end{equation}
where $\rho$ denotes the mass density. 
It is worthwhile to mention that there are no contributions from 
quadrupole moment whose components are along the line of sight 
(Asada and Kasai 2000). 

Now, a renormalisation is done in units of the Einstein ring 
angular radius defined as (for instance, Schneider et al. 1992) 
\begin{equation}
\theta_{\mbox{E}}=
\sqrt{\frac{4GMD_{\mbox{LS}}}{c^2D_{\mbox{L}}D_{\mbox{S}}}} , 
\end{equation}
so that the lens equation is simply rewritten as 
\begin{equation}
\tilde{\mbox{\boldmath $\beta$}}=
\tilde{\mbox{\boldmath $\theta$}} - \tilde{\mbox{\boldmath $\alpha$}} , 
\label{lenseq2}
\end{equation}
where 
$\tilde{\mbox{\boldmath $\beta$}}
=\mbox{\boldmath $\beta$}/\theta_{\mbox{E}}$, 
$\tilde{\mbox{\boldmath $\theta$}}
=\mbox{\boldmath $\theta$}/\theta_{\mbox{E}}$, 
and 
\begin{equation}
\tilde{\alpha}^i = 
\frac{\tilde{\theta}^i}{|\tilde{\mbox{\boldmath $\theta$}}|^2}
+ 2\sum_{j,k=1}^{2} \tilde{Q}_{jk}
\frac{\tilde{\theta}^j \tilde{\theta}^k \tilde{\theta}^i}
{|\tilde{\mbox{\boldmath $\theta$}}|^6} 
- \sum_{j=1}^{2} \tilde{Q}_{ij}\frac{\tilde{\theta}^j}
{|\tilde{\mbox{\boldmath $\theta$}}|^4} . 
\label{deflection2}
\end{equation}
Here, the renormalised quadrupole moment is defined as 
\begin{equation}
\tilde{Q}_{ij} = 
\frac{c^2D_{\mbox{S}}Q_{ij}}{2GM^2D_{\mbox{L}}D_{\mbox{LS}}} .  
\end{equation}
Eq. ($\ref{lenseq2}$) is a set of coupled equations which are 
seventh-order in $\tilde{\theta}^1$ and $\tilde{\theta}^2$, 
respectively.  
It is apparent that we can hardly treat it analytically.  

Before examining Eq. ($\ref{lenseq2}$) in detail, 
we estimate the order of magnitude of the renormalised 
quadrupole moment. We can put $|Q_{ij}| \sim \varepsilon M R^2$, 
where $\varepsilon$ denotes a distortion parameter and 
$R$ is a typical size of the lens object. 
Truncations up to quadrupole moment in multipole expansions 
are valid if the asymmetry of the system is sufficiently small, 
say $\varepsilon < 1/10$, though we don't perform any expansion 
in $\varepsilon$. 
The renormalised quadrupole moment is rewritten as 
\begin{equation}
|\tilde Q_{ij}| \sim 2\varepsilon 
\left(\frac{R}{R_{\mbox{E}}}\right)^2 , 
\end{equation}
where $R_{\mbox{E}}=D_{\mbox{L}}\theta_{\mbox{E}}$ 
is the Einstein ring radius. 
In most cases, $R$ is shorter than $R_{\mbox{E}}$. 
Thus, $|\tilde Q_{ij}|$ must be less than $2\varepsilon$, 
which is usually $1/5$ at most. 

For instance, we assume a nearby solar-type star 
at $10$pc, namely $R_{\mbox{E}}\sim 10^7$km, $R\sim 10^6$km. 
If it rotates much faster than the Sun, say with the rotational 
surface velocity on the equatorial plane $v\sim 10$km/s, 
the oblateness due to the centrifugal flattering is 
of the order of 
\begin{equation}
\varepsilon \sim 10^{-3}\left(\frac{M_{\odot}}{M}\right) 
\left(\frac{R}{10^6 \mbox{km}}\right)
\left(\frac{v}{10 \mbox{km/s}}\right)^2 . 
\end{equation}
Then, $\tilde Q_{ij}$ becomes about $10^{-5}$. 
If a light ray passes near the surface of the star, 
the leading term of the deflection angle is about $1$ arcsec. 
and the correction due to the quadrupole moment can be as large 
as $10$ microarcsec., which will be detectable in principle 
by a near future astrometry mission such as GAIA and SIM, 
though such a probability may be extremely small. 

Henceforth, we adopt a frame in which the reduced quadrupole 
moment is diagonalised as 
\begin{equation}
\tilde{Q}_{ij}=
\left(
\begin{array}{cc}
e & 0 \\
0 & -e 
\end{array}
\right) . 
\end{equation}
Without loss of generality, we can assume $e>0$. 
By denoting $\tilde{\mbox{\boldmath $\beta$}}=(a,b)$ and 
$\tilde{\mbox{\boldmath $\theta$}}
=(x, y)=(r \cos\phi, r \sin\phi)$ for $r \geq 0$, 
it is convenient to rewrite Eqs. ($\ref{lenseq2}$) and 
($\ref{deflection2}$) into a matrix form 
\begin{equation}
\left(
\begin{array}{c}
a \\ b 
\end{array}
\right)
=
\left(
\begin{array}{cc}
\cos\phi & \cos3\phi \\
\sin\phi & \sin3\phi
\end{array}
\right)
\left(
\begin{array}{c}
f \\ g 
\end{array}
\right) , 
\label{lenseqmatrix1}
\end{equation}
where 
\begin{eqnarray}
f&=&r-\frac1r , \\
g&=&-\frac{e}{r^3} . 
\end{eqnarray}
Here, $f$ and $g$ depend on the mass and quadrupole moment, 
respectively, where we should remember that our renormalisation 
was done in units of the Einstein ring radius so that the mass 
does not appear explicitly. 

\subsection{Sources on the symmetry axes}
First of all, we consider a source located at the origin 
$(a,b)=(0, 0)$, namely exactly behind the lens object. 
For $b=0$, the second line of Eq. ($\ref{lenseqmatrix1}$) 
means (1) $\sin\phi=0$ or (2) $f+g(3-4\sin^2\phi)=0$. 
The case (1) means $y=0$, consequently $r^2=x^2$. 
Then, the first line of Eq. ($\ref{lenseqmatrix1}$) 
becomes 
\begin{equation}
x^4-x^2-e=0 , 
\end{equation}
where we used $r\neq 0$. 
This equation is solved immediately as 
\begin{equation}
x=\pm\sqrt{\frac{1+\sqrt{1+4e}}{2}} , 
\label{case1sol}
\end{equation}
where we used $\sqrt{1+4e}>1$. 
Next, we consider the case (2), whose condition 
is rewritten as 
\begin{equation}
f+4g\cos^2\phi=g . 
\label{case2}
\end{equation}
By substituting this into the first line of 
Eq. ($\ref{lenseqmatrix1}$), we obtain $g\cos\phi=0$. 
Because of $g\neq 0$, we find $\cos\phi=0$, which means 
$x=0$ and $r^2=y^2$. 
Then, Eq. ($\ref{case2}$) becomes simply $f=g$, 
which is rewritten as $y^4-y^2+e=0$. 
This is solved as 
\begin{equation}
y=\pm\sqrt{\frac{1\pm\sqrt{1-4e}}{2}} ,  
\label{case2sol}
\end{equation}
which are real if and only if $e\leq 1/4$. 
Hence, we find two images given by Eq. ($\ref{case1sol}$) 
on $x$-axis, and four images by Eq. ($\ref{case2sol}$) 
on $y$-axis if $e\leq 1/4$, 
or otherwise only the first two images on $x$-axis. 

Now, let us consider a case that a source is located 
on a symmetry axis but off the origin. 
For simplicity, we assume $b=0$ and $a\neq 0$. 
The case of $a=0$ and $b\neq 0$ is obtained simply 
by transforming $e\to -e$ and $x\leftrightarrow y$ as discussed later. 
For $b=0$, the second line of Eq. ($\ref{lenseqmatrix1}$) 
means (a) $\sin\phi=0$ or (b) $f+g(3-4\sin^2\phi)=0$. 
The case (a) means $y=0$ and $r^2=x^2$. 
Then, the first line of Eq. ($\ref{lenseqmatrix1}$) 
becomes 
\begin{equation}
u(x)\equiv 
x^4-ax^3-x^2-e=0 , 
\label{casea}
\end{equation}
where we used $r\neq 0$. 
The discriminant for this fourth-order polynomial 
(van der Waerden 1966) is computed as 
\begin{equation}
D_4=-e (16+128e+256e^2+4a^2+144a^2e+27a^4e) , 
\end{equation}
which is negative. 
This implies that Eq. ($\ref{casea}$) has 
only two real roots.  
Furthermore, we find $u(0)=-e<0$. 
The corresponding two images are located on $x$-axis, 
one positive and one negative each. 

The case of $a=0$ and $b\neq 0$ is more complicated. 
The transformation mentioned above 
gives us 
\begin{equation}
v(y)\equiv 
y^4-by^3-y^2+e=0 , 
\end{equation}
which is the second line of Eq. ($\ref{lenseqmatrix1}$). 
For this equation, the discriminant becomes 
\begin{equation}
D_4=e (16-128e+256e^2+4b^2-144b^2e-27b^4e) . 
\end{equation}
$D_4=0$ has two roots
\begin{equation}
b^2=b_{\pm}^2 , 
\end{equation}
where we defined 
\begin{eqnarray}
b_{\pm}^2=\frac{2}{3^3e}
[
(1-2^23^2e)
\pm (1+2^23e)^{3/2} ] . 
\end{eqnarray}
One can show that $b_{-}^2 < 0$ and $b_{+}^2 > 0$. 
Hence, for $b^2 > b_{+}^2$, we find $D_4 < 0$, which 
implies two real roots, namely two on-axis images. 
On the other hand, for $b^2 < b_{+}^2$, we find $D_4 > 0$, 
which implies either four or no real roots. Further investigations 
are thus needed. 
We use the derivative of $v(y)$ as 
\begin{equation}
v^{\prime}(y)=y(4y^2-3by-2) . 
\end{equation}
The solutions of $v^{\prime}(y)=0$ are 
$y=0, y_{\pm}$, where we defined 
\begin{equation}
y_{\pm}=\frac{1}{2^3}(3b\pm\sqrt{3^2b^2+2^5}) . 
\end{equation}
If $v(y_{+}) < 0$ and $v(y_{-}) < 0$, four real roots exist, 
while no real roots for $v(y_{+}) > 0$ and $v(y_{-}) > 0$. 
In the case of $D_4 > 0$, the former condition is equivalent to 
$v(y_{+}) + v(y_{-}) < 0$, which is solved as 
\begin{equation}
b^2 > b_{\times}^2 
\equiv \frac{2^3}{3}(-1+3^{-1/2}\sqrt{1+2^3e}) . 
\end{equation}
The term $b_{\times}^2$ is positive if and only if $e > 1/4$. 
For $e > 1/4$, one can show $b_{\times}^2 > b_{+}^2$. 
If $e > 1/4$, we thus find no images for $b^2<b_{+}^2$ 
and two for for $b^2>b_{+}^2$, 
while for $e < 1/4$, 
four images exist if $b^2 < b_{+}^2$, 
and two images if $b^2 > b_{+}^2$. 

We consider the case (b), whose condition 
is rewritten as 
\begin{equation}
f+4g\cos^2\phi=g . 
\label{caseb}
\end{equation}
By substituting this into the first line of 
Eq. ($\ref{lenseqmatrix1}$), we obtain $a=-2g\cos\phi$. 
Because of $g\neq 0$, we find 
\begin{equation}
\cos\phi=\frac{ar^3}{2e} . 
\label{caseb2}
\end{equation}
Substitution of this into Eq. ($\ref{caseb}$) gives us 
a cubic equation for $r^2$ as 
\begin{equation}
a^2r^6-er^4+er^2-e^2=0 . 
\label{caseb3}
\end{equation}
Because of the symmetry with respect to x-axis, 
images may appear at $(r\cos\phi, \pm r\sqrt{1-\cos\phi^2})$, 
where $r$ is a solution of Eq. ($\ref{caseb3}$) 
and $\cos\phi$ is obtained by substituting the solution 
for $r$ into Eq. ($\ref{caseb2}$). 

Here, we should note a constraint on $r$, 
which comes from Eq. ($\ref{caseb2}$) 
for $|\cos\phi|\leq 1$. 
The constraint is expressed as 
\begin{equation}
r\leq \left|\frac{2e}{a}\right|^{1/3} . 
\label{constraint}
\end{equation}

One can investigate the number of off-axis images 
according to Eqs. ($\ref{caseb3}$) and ($\ref{constraint}$). 
The detailed discussion is given in the Appendix A. 
It shows that the maximum number of off-axis images 
is four if $0<e<1/4$, six if $1/4<e<1/3$, and two if $e>1/3$. 
Similar results in a case of $a=0$ and $b\neq 0$ are 
given in the Appendix B.

\subsection{Off-axis sources} 
Here, let us consider off-axis sources, $a\neq 0$ and $b\neq 0$. 
Then, Eq. ($\ref{lenseqmatrix1}$) implies $\sin\phi\neq 0$ 
and $\cos\neq 0$, which are equivalent to $\sin2\phi \neq 0$. 
By using an inverse matrix, which always exists if 
$\sin2\phi \neq 0$, Eq. ($\ref{lenseqmatrix1}$) is rewritten as 
\begin{eqnarray}
\left(
\begin{array}{c}
f \\ g 
\end{array}
\right)
&
=
&
\frac{1}{\sin2\phi}
\left(
\begin{array}{cc}
\sin3\phi & -\cos3\phi \\
-\sin\phi & \cos\phi
\end{array}
\right)
\left(
\begin{array}{c}
a \\ b 
\end{array}
\right) 
\nonumber\\
&
\equiv
& 
\left(
\begin{array}{c}
F \\ G 
\end{array}
\right) .  
\label{lenseqmatrix2}
\end{eqnarray}
Thus $r$-dependent parts are separated from $\phi$-dependent ones. 
A key is the first line of Eq. ($\ref{lenseqmatrix2}$), which 
is re-expressed as $r^2=F r+1$. 
By using this recursively, we obtain $r^3=(F^2+1)r+F$.  
Substituting this into $r^3$ in the left hand side of 
the second line of Eq. ($\ref{lenseqmatrix2}$), we obtain
\begin{equation}
r=-\frac{FG+e}{G(F^2+1)} , 
\label{r}
\end{equation}
whose right-hand side depends only on $\phi$. 
This equation implies that $G$ does not vanish for $e \neq 0$ 
because $r$ is finite. 
We should note that $r$ cannot be expressed 
as a function only of $\cot\phi$ but also of $\cos\phi$. 

Now, we rewrite $F$ and $G$ as 
\begin{eqnarray}
F&=&\frac{F_0}{2(1+u^2)\cos\phi} , 
\label{F}\\
G&=&\frac{b u - a}{2\cos\phi} , 
\label{G}
\end{eqnarray}
where 
\begin{eqnarray}
u&=&\cot\phi , \\
F_0&=&a(3u^2-1)-bu(u^2-3) . 
\end{eqnarray}
Thus, by using Eqs. ($\ref{r}$), ($\ref{F}$) and ($\ref{G}$), 
we can express $(x, y)$ as a function only of $u$, 
\begin{eqnarray}
x&=&
-\left(\frac{F_0}{2(1+u^2)}+\frac{2e u^2}{(bu-a)(1+u^2)}\right)
\nonumber\\
&&
\times\left({\frac{F_0^2}{4u^2(1+u^2)}+1}\right)^{-1} , \\
y&=&
-\left(\frac{F_0}{2u(1+u^2)}+\frac{2e u}{(bu-a)(1+u^2)}\right)
\nonumber\\
&&
\times\left({\frac{F_0^2}{4u^2(1+u^2)}+1}\right)^{-1} . 
\end{eqnarray} 
This fact implies that there must exist a master equation 
determining $u$. 
Substituting Eq. ($\ref{r}$) into $r$ in the first line of 
Eq. ($\ref{lenseqmatrix2}$), we obtain 
a tenth-order equation for $u=\cot\phi$
\begin{equation}
\sum_{k=0}^{10} c_k u^k = 0 , 
\label{eq-u}
\end{equation}
where each coefficient is a polynomial in $a$, $b$ and $e$ as 
\begin{eqnarray}
c_{10}&=&b^4e , \\
c_9&=&-10ab^3e , \\
c_8&=&4b^2+3b^2e(4+12a^2-3b^2) , \\
c_7&=&-8ab-6abe(8+9a^2-11b^2) , \\
c_6&=&4a^2+8b^2+3e(12a^2+9a^4-8b^2-52a^2b^2+9b^4)
\nonumber\\
&&-16e^2 , \\
c_5&=&-16ab+126abe(a^2-b^2) , \\
c_4&=&8a^2+4b^2+3e(8a^2-9a^4-12b^2+52a^2b^2-9b^4)
\nonumber\\
&&-16e^2 , \\
c_3&=&-8ab+6abe(8-11a^2+9b^2) , \\
c_2&=&4a^2-3a^2e(4-3a^2+12b^2) , \\
c_1&=&10a^3be , \\
c_0&=&-a^4e . 
\end{eqnarray} 

As seen from Eq. ($\ref{eq-u}$), $u=\cot\phi$ neither vanishes 
nor diverges because $c_{10}\neq 0$ and $c_0\neq 0$ 
for $a$, $b$ and $e\neq 0$. 
The number of images is constant unless a source crosses 
the caustics. 
Therefore, the maximum number of real solutions for 
Eq. ($\ref{eq-u}$) is eight, which equals that for an on-axis source 
as shown in the preceding subsection. 
One can see also that there is $c_i\leftrightarrow c_{10-i}$ 
correspondence with $a\leftrightarrow b$ and $e\leftrightarrow -e$ 
transformation. This is taken as a justification for 
the statement regarding the relation 
between $(a=0, b\neq 0)$ and $(a\neq 0, b=0)$.

\section{Discriminant and Caustics}
As well-known in algebra, a discriminant for a real polynomial 
tells us where the number of real roots changes 
(for instance, van der Waerden 1966). 
A discriminant for Eq. ($\ref{eq-u}$) is obtained for instance 
by using Mathematica (Wolfram 2000) as 
\begin{equation}
D_{10}=2^{42} a^{12}b^{12}e^6 W^2 K , 
\end{equation}
where  
\begin{eqnarray}
W&=&(a^2+b^2)^2+(a^2+b^2)^3-2e(a^2-b^2)
\nonumber\\
&&
+e^2[1-(a^2+b^2)+3(a^2+b^2)^2] 
\nonumber\\
&&-e^4[2-3(a^2+b^2)]
+e^6 ,
\label{W}
\end{eqnarray}
and $K$ is a lengthy polynomial of 494 terms as 
\begin{equation}
K=2^2 3^{16} a^{22} e^5(2+18e+27b^2e)+\cdots . 
\label{caustics}
\end{equation}
This discriminant $D_{10}$ gives us a condition for changes 
in the number of real roots for Eq. ($\ref{eq-u}$), 
corresponding to changes in the number of images. 

It should be noted that $W$ never vanishes as shown below, 
which means that $r$ is uniquely determined by Eq. ($\ref{r}$). 
This is in contrast to a case of a binary gravitational lens, 
for which a squared term such as $W^2$ can vanish so that  
a function of $r$ can take a form of ``$0/0$'' 
for particular cases (Asada et al. 2002).  
By paying attention to an asymmetric term proportional 
to $a^2-b^2$ in Eq. ($\ref{W}$), we rewrite $W$ as 
\begin{eqnarray}
W&=&(a^2+b^2-e+e^2)^2 [a^2+b^2+(1+e)^2] + 4b^2e , 
\end{eqnarray} 
which shows that $W$ is always positive 
for $b\neq 0$ and $e>0$. 
The caustics are thus expressed simply as $K=0$.

\section{Critical Curves} 
In the preceding section, the explicit expression for 
the caustics has been obtained. If we map the expression 
onto the lens plane by the lens equation, we can obtain 
the corresponding expression for the critical curves. 
In practice, however, it is a quite difficult task. 
Hence, let us study the critical curves in a different way. 
By a straightforward calculation of the vanishing Jacobian 
of the lens mapping by Eq. ($\ref{lenseqmatrix1}$), 
\begin{equation}
\frac{\partial(a, b)}{\partial(x, y)} = 
\frac{1}{r} \frac{\partial(a, b)}{\partial(r, \phi)} = 0 , 
\end{equation} 
we obtain a simple expression for the critical curves as 
\begin{equation}
r^8=r^4+9e^2+6er^2\cos2\phi . 
\label{critical}
\end{equation}
This curve is classified topologically into two as follows. 
Because of $|\cos2\phi|\leq 1$, Eq. ($\ref{critical}$) 
gives a constraint on $r$ 
\begin{equation}
|r^8-r^4-9e^2|\leq 6er^2 , 
\end{equation} 
which is equivalent to 
\begin{equation}
r^4\leq |r^2+3e| , 
\label{inequality1}
\end{equation}
and 
\begin{equation}
r^4\geq |r^2-3e| , 
\label{inequality2}
\end{equation}
where we used $r^4\geq 0$. 
By solving carefully Eqs. ($\ref{inequality1}$) and 
($\ref{inequality2}$), we find for $e\geq 1/12$ 
\begin{equation}
\frac{-1+\sqrt{1+12e}}{2}\leq r^2\leq \frac{1+\sqrt{1+12e}}{2} , 
\label{inequality11}
\end{equation}
and for $e < 1/12$, 
\begin{eqnarray}
\frac{-1+\sqrt{1+12e}}{2}\leq &r^2& 
\leq \frac{1-\sqrt{1-12e}}{2} , 
\label{inequality21}
\end{eqnarray} 
or 
\begin{eqnarray}
\frac{1+\sqrt{1-12e}}{2}\leq &r^2&
\leq \frac{1+\sqrt{1+12e}}{2} . 
\label{inequality22}
\end{eqnarray} 
This implies a topological feature of the critical curves: 
The curves are simply connected if $e\geq 1/12$, 
or otherwise they are not. 

When Eq. ($\ref{critical}$) is treated as a $\phi$-parameter 
representation of the critical curves, we must solve it 
as a fourth-order equation for $r^2$, so that $r^2$ can be
multi-valued. As a consequence of $r\geq 0$, furthermore, 
some domains of the parameter $\phi\in [0, 2\pi)$ 
can be excluded. 
It seems thus much simpler to consider Eq. ($\ref{critical}$) 
as a representation with a parameter $r$ which is allowed 
by Eq. ($\ref{inequality11}$) for $e\geq 1/12$ 
or by Eqs. ($\ref{inequality21}$) and ($\ref{inequality22}$) 
for $e < 1/12$. 
The representation becomes 
\begin{equation}
(x, y)=(\pm r\sqrt{\frac{1+h}{2}}, \pm r\sqrt{\frac{1-h}{2}}) , 
\label{critical-r}
\end{equation}
where we defined 
\begin{equation}
h=\frac{r^8-r^4-9e^2}{6er^2} . 
\end{equation}

In addition, substitution of this parametric representation 
into the right-hand side of Eq.($\ref{lenseqmatrix1}$) gives 
us a representation of the caustics with the same parameter as 
\begin{eqnarray}
a(r)&=&\pm[f+g(2h-1)]\sqrt{\frac{1+h}{2}} , 
\label{caustics-a}
\\
b(r)&=&\pm[f+g(2h+1)]\sqrt{\frac{1-h}{2}} . 
\label{caustics-b}
\end{eqnarray}

\begin{figure}
\includegraphics[width=8cm]{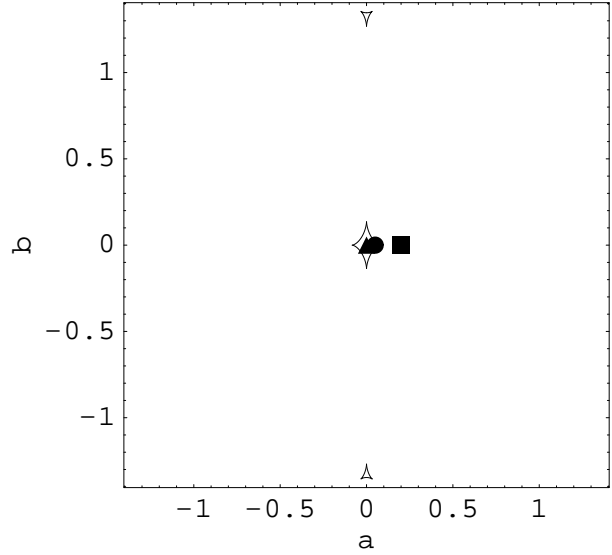}
\caption{
Caustics for a lens with quadrupole moment $e=0.05$. 
Sources are located at $(0, 0)$, $(0.05, 0)$ and $(0.2, 0)$, 
denoted by the triangle, filled disk and square, respectively. 
}
\label{Fig1}
\end{figure}

\begin{figure}
\includegraphics[width=8cm]{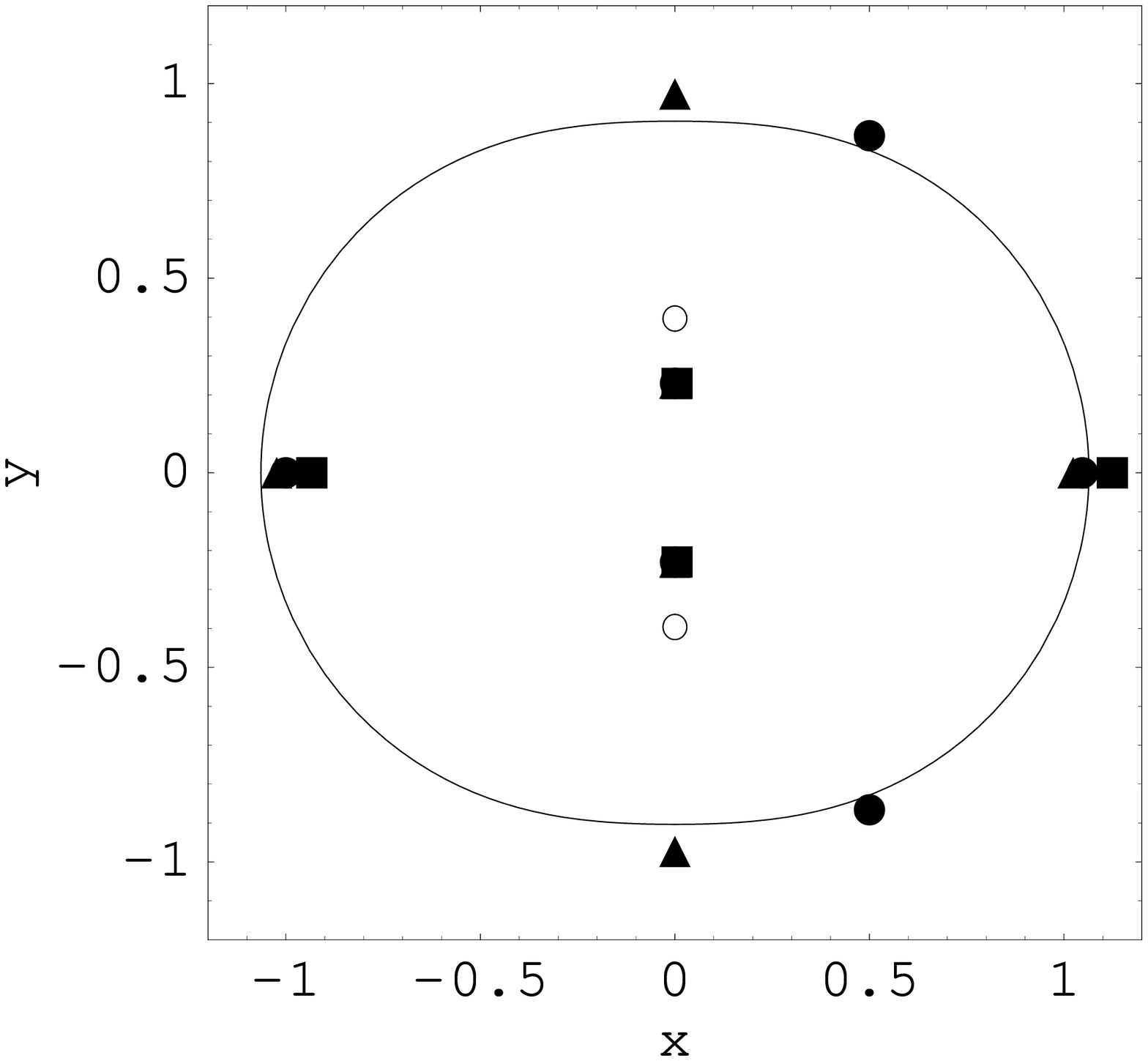}
\caption{
Critical curves for a lens with quadrupole moment $e=0.05$. 
The images correspond to the sources in Fig. 1. 
Triangles, Filled disks and squares are overlapped around 
$(0, \pm0.2)$.}
\label{Fig2}
\end{figure}

\begin{figure}
\includegraphics[width=8cm]{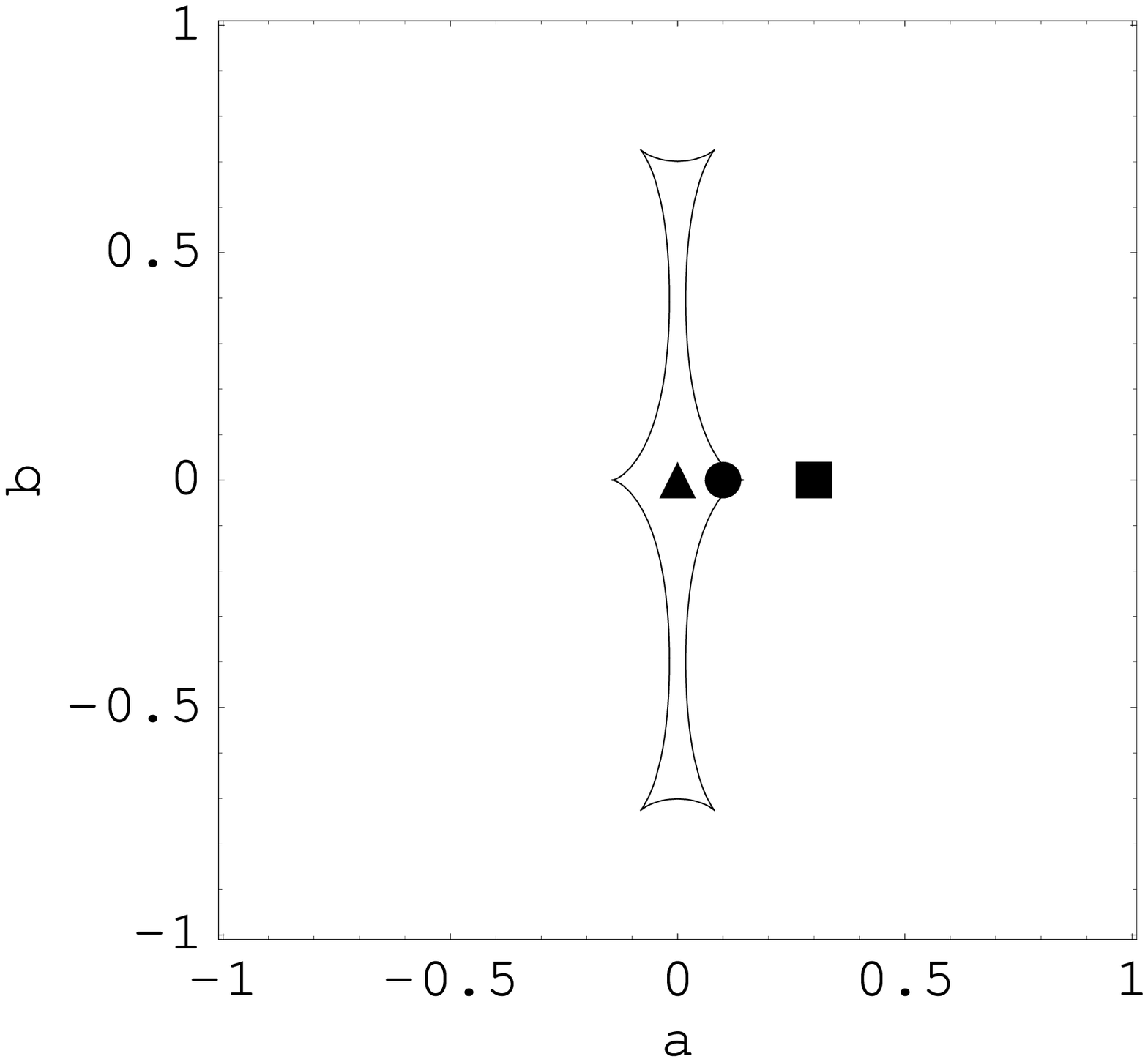}
\caption{
Caustics for a lens with quadrupole moment $e=0.1$. 
Sources are located at $(0, 0)$, $(0.1, 0)$ and $(0.3, 0)$, 
denoted by the triangle, filled disk and square, respectively. 
}
\label{Fig3}
\end{figure}

\begin{figure}
\includegraphics[width=8cm]{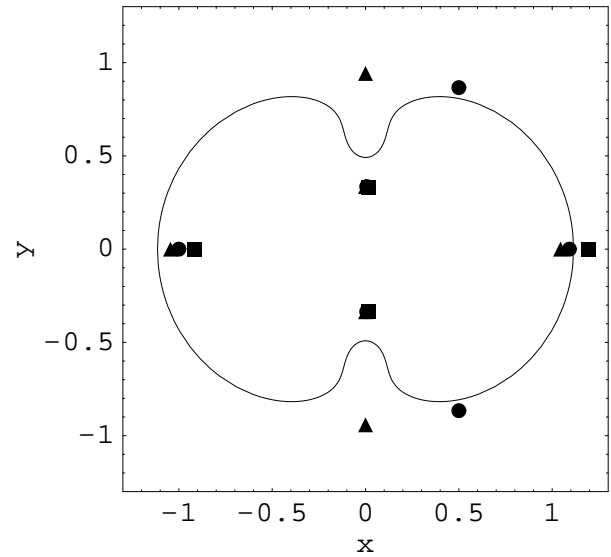}
\caption{
Critical curves for a lens with quadrupole moment $e=0.1$. 
The images correspond to the sources in Fig. 3. 
Triangles, Filled disks and squares are overlapped around 
$(0, \pm0.3)$.}
\label{Fig4}
\end{figure}

As illustrations, the caustics and critical curves 
for $e=0.05<1/12$ are given by Figs. 1 and 2, respectively. 
Actually, there are three closed loops of critical curves. 
A case of $e=0.1>1/12$ is shown by Figs. 3 and 4.

\section{Conclusion}
We have re-examined the lens equation for a gravitational 
lens due to a deformed star modeled by using quadrupole moment.  
First, we reduce the lens equation to a single real tenth-order 
polynomial. 
Consequently, an analytic expression for the caustics 
is given by Eq. ($\ref{caustics}$) as $K=0$, 
though it is too lengthy to handle.  
The critical curves are simply expressed as 
Eq. ($\ref{critical}$) or Eq. ($\ref{critical-r}$). 
The parametric representation of the caustics is given 
by Eqs. ($\ref{caustics-a}$) and ($\ref{caustics-b}$). 
Hence, the present result must be helpful for understanding 
the gravitational lensing due to a deformed star. 
In particular, it gives a nice approximation to a lensing study 
based on a realistic numerical simulation of a compact object. 

For instance, the image position is unstable for a source 
close to the caustics, so that careful numerical computations 
are needed. 
The present approach will make it easy to study such a case. 
This is a future subject.

\section*{ACKNOWLEDGMENTS}
The author would like to thank the anonymous referee for 
invaluable comments especially about the classification 
of the number of images for on-axis sources. 
He would like to thank E. Berti for fruitful conversations. 
He would like to thank L. Blanchet for hospitality at the Institute 
for Astrophysics at Paris. 
This work was supported by a fellowship for visiting scholars from 
the Ministry of Education of Japan.

\appendix

\section[]{Classifications of off-axis images I.}
Let us investigate the number of off-axis images 
according to Eqs. ($\ref{caseb3}$) and ($\ref{constraint}$) 
in a case of $a\neq 0$ and $b=0$. 
They are rewritten as  
\begin{equation}
g(p)\equiv a^2p^3-ep^2+ep-e^2=0 , 
\label{g}
\end{equation}
\begin{equation}
p\leq \left|\frac{2e}{a}\right|^{2/3} \equiv p_0 ,  
\label{gconstraint} 
\end{equation}
where we defined $p=r^2$. 
The discriminant of Eq. ($\ref{g}$) is 
\begin{equation}
D_3=-e^3 [3^3 e a^4+(2^2-2\cdot3^2e)a^2-(e-2^2e^2)] . 
\end{equation}
$D_3=0$ has two roots as 
\begin{equation}
a^2=A_{\pm} , 
\end{equation}
where we defined 
\begin{equation}
A_{\pm}=\frac{1}{3^3e}[(3^2e-2) \pm 2(1-3e)^{3/2}] . 
\end{equation}
We find $A_{-} > 0$ if $e > 1/4$, and otherwise $A_{-} < 0$. 
We find $g(p_0)$ as 
\begin{equation}
g(p_0)=\frac{e^{5/3}}{a^{4/3}}
\left(
3e^{1/3}a^{4/3}-2^{4/3}e^{2/3}+2^{2/3}a^{2/3} 
\right) . 
\end{equation} 
The condition that $g(p_0) > 0$ is 
\begin{equation}
a^2 > A_0 
\equiv 
\frac{2}{3^3e}[-(1+3^2e)+(1+3e)\sqrt{1+2^23e}] , 
\end{equation}
where we used the positivity of $a^2$.  

In order to analyse the number of the valid solutions 
satisfying Eq. ($\ref{gconstraint}$), one needs to investigate  
the signature not only of $g(p_0)$ but also of $g'(p_0)$. 
In a case of $g(p_0)<0$ and $g'(p_0)>0$, for instance, 
one can show that there exists a root $p > p_0$, 
which corresponds to an unphysical solution $r$ 
violating the condition $(\ref{constraint})$. 
The condition that $g'(p_0) < 0$ is 
\begin{equation}
a^2 < A_1 
\equiv 
\frac{1}{2^53^3e}[-(1+2^33^2e)+(1+2^33e)\sqrt{1+2^53e}] , 
\end{equation}
where we used the positivity of $a^2$.  

The number of the valid solutions are found by figuring out 
the relative order among $a^2$, $A_{\pm}$, $A_0$ and $A_1$, 
which depends on the value of $e$. Straightforward but tedious 
computations give the following result. 

\noindent
(1) $0<e<\frac14$ \\
(1a) $A_{-}<0<a^2<A_0<A_{+}<A_1$ \\
There are three positive roots. Only two of them are valid and 
thus four off-axis images exist. \\
(1b) $A_{-}<A_0<a^2<A_{+}<A_1$ \\
There are three positive roots. Only one of them is valid and 
thus two off-axis images exist. \\
(1c) $A_{-}<A_0<A_{+}<a^2<A_1$ \\
There is only one positive root, which is still valid. 
Thus two off-axis images exist. \\
(1d) $A_{-}<A_0<A_{+}<A_1<a^2$ \\
There is only one positive root, which is still valid.  
Thus two off-axis images exist. 

\noindent
(2) $\frac14<e<\frac{8}{25}$ \\
The relative order between $A_{-}$ and $A_1$ is not unique 
in this case.\\ 
(2a) $0<a^2<A_{-}, A_1<A_0<A_{+}$ \\
There is only one positive root, which is not valid.  
Thus no off-axis images exist. \\
(2b) $0<A_{-}<a^2<A_1<A_0<A_{+}$ \\
There are three positive roots. Only two of them are valid and 
thus four off-axis images exist. \\
(2c) $0<A_1<a^2<A_{-}<A_0<A_{+}$ \\
There is only one root, which is not valid. 
Thus no off-axis images exist.\\
(2d) $0<A_{-}, A_1<a^2<A_0<A_{+}$ \\
There are three positive roots. Only two of them are valid and  
thus four off-axis images exist. \\
(2e) $0<A_{-}, A_1<A_0<a^2<A_{+}$ \\
There are three positive roots, all of which are still valid.  
Thus six off-axis images exist. \\
(2f) $0<A_{-}, A_1<A_0<A_{+}<a^2$ \\
There are only one positive root, which is still valid.  
Thus two off-axis images exist. 

\noindent
(3) $\frac{8}{25}<e<\frac13$ \\
(3a) $0<a^2<A_1<A_0<A_{-}<A_{+}$ \\
There is only one positive root, which is not valid.  
Thus no off-axis images exist. \\
(3b) $0<A_1<a^2<A_0<A_{-}<A_{+}$ \\
There is only one positive root, which is not valid. 
Thus no off-axis images exist. \\
(3c) $0<A_1<A_0<a^2<A_{-}<A_{+}$ \\
There is only one positive root, which is still valid.   
Thus two off-axis images exist. \\
(3d) $0<A_1<A_0<A_{-}<a^2<A_{+}$ \\
There are three positive roots, all of which are still valid.  
Thus six off-axis images exist. \\
(3e) $0<A_1<A_0<A_{-}<A_{+}<a^2$ \\
There is only one positive root, which is still valid.  
Thus two off-axis images exist. 

\noindent
(4) $\frac13<e$ \\
(4a) $0<a^2<A_1<A_0$ \\
There is only one positive root, which is not valid.  
Thus no off-axis images exist. \\
(4b) $0<A_1<a^2<A_0$ \\
There is only one positive root, which is not valid.  
Thus no off-axis images exist. \\
(4c) $0<A_1<A_0<a^2$ \\
There is only one positive root, which is still valid.  
Thus two off-axis images exist. 

\section[]{Classifications of off-axis images II.}
We classify the number of off-axis images 
in a case of $a=0$ and $b\neq 0$. 
The transformation of $a\leftrightarrow b$ and 
$e\leftrightarrow -e$ changes Eq. $(\ref{caseb3})$ as 
\begin{equation}
b^2r^6+er^4-er^2-e^2=0 . 
\label{cubic-b}
\end{equation}
The constraint by Eq. $(\ref{constraint})$ is also 
transformed into 
\begin{equation}
r\leq \left|\frac{2e}{b}\right|^{1/3} . 
\label{constraint-b}
\end{equation}

Let us investigate the number of off-axis images 
according to Eqs. ($\ref{cubic-b}$) and ($\ref{constraint-b}$) 
in a case of $a=0$ and $b\neq 0$. 
They are rewritten as  
\begin{equation}
h(q)\equiv b^2q^3+eq^2-eq-e^2=0 , 
\label{h}
\end{equation}
\begin{equation}
q\leq \left|\frac{2e}{b}\right|^{2/3} \equiv q_0 ,  
\label{hconstraint} 
\end{equation}
where we defined $q=r^2$. 

One can show that the positive root is always only one 
because $h(0)<0$ and $h^{\prime}(0)<0$.  
This situation is much simpler than that of $a\neq 0$ and $b=0$. 
Namely, one can show that the valid solution is still one 
if $h(q_0)>0$, while we have no valid solution if $h(q_0)<0$. 
We find $h(q_0)>0$ always holds if $e>1/12$. 
If $e<1/12$, $h(q_0)>0$ is equivalent to 
$b^2<B_{-}$ or $b^2>B_{+}$, where we defined 
\begin{equation}
B_{\pm}=\frac{2}{3^3e}[(1-3^2e)\pm (1-3e)\sqrt{1-2^2\cdot 3e}] . 
\end{equation}
As a result, the number of the off-axis images is classified as 
follows.  

\noindent
(1) $e<\frac{1}{12}$ \\
(1a) $0<b^2<B_{-}$ \\
There is only one positive root, which is still valid.  
Thus two off-axis images exist. \\
(1b) $B_{-}<b^2<B_{+}$ \\
There is only one positive root, which is not valid.  
Thus no off-axis images exist. \\
(1c) $B_{+}<b^2$ \\
There is only one positive root, which is still valid.  
Thus two off-axis images exist. 

\noindent
(2) $e>\frac{1}{12}$ \\
(2a) $0<b^2$ \\
There is only one positive root, which is still valid.  
Thus two off-axis images exist.

\bsp 


\end{document}